\documentclass[showpacs,preprintnumbers,amsmath,amssymb]{revtex4}
\usepackage{graphicx}
\usepackage{amssymb}

\begin{document}

\title{Quantum Criticality in doped CePd$_{1-x}$Rh$_x$ Ferromagnet}

\author{J.G. Sereni$^1$, T. Westerkamp$^2$, R. K\"uchler$^2$, N. Caroca-Canales$^2$, P. Gegenwart$^2$, C. Geibel$^2$}
\address{$^1$ Lab. Bajas Temperaturas, Centro At\'omico Bariloche, 8400 S.C. Bariloche, Argentina\\
$^2$ Max-Planck Institute for Chemical Physics of Solids, D-01187 Dresden, Germany}

\date{\today}-

\begin{abstract}

{CePd$_{1-x}$Rh$_x$ alloys exhibit a continuous evolution from ferromagnetism ($T_C=6.5$\,K) at $x = 0$ to a mixed valence (MV) state at $x = 1$. We have
performed a detailed investigation on the suppression of the ferromagnetic (F) phase in this
alloy using dc-($\chi_{dc}$) and ac-susceptibility ($\chi_{ac}$), specific
heat ($C_m$), resistivity ($\rho$) and thermal expansion ($\beta$) techniques.
Our results show a continuous decrease of $T_C(x)$ with negative curvature
down to $T_C$ = 3\,K at $x^*=0.65$, where a positive curvature takes over.
Beyond $x^*$, a cusp in $\chi_{ac}$ is traced down to $T_C^* =25$\,mK at
$x=0.87$, locating the critical concentration between $x=0.87$ and $0.90$. The
quantum criticality of this region is recognized by the $-log(T/T_0)$
dependence of $C_{m}/T$, which transforms into a $T^{-q}$ ($q\approx 0.5$) one
at $x=0.87$. At high temperature, this system shows the onset of valence
instability revealed by a deviation from Vegard's law (at $x_V \approx 0.75$)
and increasing hybridization effects on high temperature $\chi_{dc}$ and
$\rho(T)$. Coincidentally, a Fermi liquid contribution to the specific heat 
arises from the MV component, which becomes dominant at the CeRh
limit. In contrast to antiferromagnetic systems, no $C_{m}/T$ flattening is observed for
$x>x_{cr}$ rather the mentioned power law divergence, which coincides with a change of sign of 
$\beta(T)$. 
The coexistence of F and MV
components and the sudden changes in the $T$ dependencies are discussed in the
context of randomly distributed magnetic and Kondo couplings.}

\end{abstract}

\maketitle

\section{Introduction}
The physics related to magnetic critical points at very low
temperature, where quantum fluctuations compete with classical
thermal fluctuations, is a topic of increasing attraction. The
wealthy spectrum of recently discovered new properties has
triggered intense experimental and theoretical activity, involving
a large number of magnetic phase diagrams.\cite{SCES} The magnetic
phase boundaries of those systems are tuned to different classes
of critical points,\cite{2Steward,1Voijta} where the long range
order is suppressed by applying pressure or alloying. Since most
of the current investigations are devoted to antiferromagnetic
(AF) systems, suitable candidates for the study of ferromagnetic
(F) systems remain scarce.\cite{3Evans} Not only the number of
available exemplary AF or F systems makes the difference between
them, rather intrinsic physical properties like those observed in
3$d$ compounds \cite{3bPfei} or the proposed by theoretical
models.\cite{5Patrick} However, up to date the great amount of AF
systems exhibiting coexistence of magnetic order and Kondo effect
with respect to the number of F ones remains a puzzling problem.
This asymmetry cannot be simply explained by a distribution of
$Ce-Ce$ spacings in their respective lattices, where inter-site
magnetic interactions ($J_R$) act mediated by the RKKY mechanism.
In spite of that, it is observed that F Ce binary compounds only
appear within a narrow $Ce-Ce$ spacing range: 3.7$< d_{Ce-Ce}
<$4.1$\AA$.\cite{4Ser91} Furthermore, in coincidence with the fact
that Kondo effect is related to an AF spin-electron-coupling
parameter ($J_K$), none of the known F Ce compounds show
conclusive indications of significant hybridization effects, which
even excludes heavy fermion behavior. Consequently, the F- order
is generally taken as hallmark of trivalent Ce$^{3+}$ ground state
\cite{3Evans,4Ser91} when it reaches the expected entropy
($S_m=R\ln 2$) at $T\approx T_C$.

The few known F Ce phase diagrams studied as a function of
pressure ($p$) as control parameter on Ce-binary compounds also
show some peculiarities. For example, no superconductivity was
detected under pressure, whereas the $T_C(p)$ phase boundary show
the characteristic maximum described by Doniach-Lavagna
model.\cite{6Doniach} Under alloying, other intrinsic differences
between F and AF systems are observed concerning their respective
``final'' (non-magnetic) ground states (GS). Whereas the former
systematically exhibit a {\it mixed-valence} (MV) state, the
latter show {\it heavy-fermion} (HF) behavior.\cite{8Ser95} Such a
difference indicates that a strong hybridization (even including
charge fluctuations) is required to overcome a F-GS, whereas in AF
spin fluctuations (with $T_K\simeq 10$\thinspace K) are enough to
screen Ce magnetic moments.\cite{8Ser95,9Coq96} Concerning ternary
compounds like CeRu$_2$Ge$_2$ \cite{Sulow} and CePd$_2$Al$_2$Ga,
\cite{Eichler} both undergo a transition to AF phases before to
reach the critical pressure.

The distinctive features between F- and AF-systems suggest
intrinsic differences in their critical points. In fact, theory
predicts that magnetic, thermal and transport properties differ
between them.\cite{5Patrick} In pure compounds (tuned by pressure)
the F-GS is believed to end in a first order transition,
disappearing at a classical critical point at finite temperature.
On the contrary, some disorder introduced by alloying could smear
out the first order phase transition, resulting in a continuous
disappearance of magnetic order. \cite{Vojta03} These predictions
make F-alloys particularly interesting for the study of F-CP.

Among F-Ce compounds driven to non-magnetic state by doping
Ce-ligands, the binary CePd$_{1-x}$Rh$_x$ is one of the most
suitable for this study. This system evolves from a $F-Ce^{3+}$
state with $T_C=6.6$\,K to a non-magnetic mixed-valence one, with
susceptibility ($\chi_{dc}$) and electrical resistivity ($\rho$)
maxima at $T\approx 280K$.\cite{11CePdRh} Its cell volume
decreases continuously with $x$ showing a deviation from Vegard's
law around $x_V= 0.75$, while $L_{III}$-XAS measurements
\cite{11CePdRh} indicate a significant decrease of the 4$f$ state
occupancy beyond that concentration. Early studies
\cite{12CePdNiRh} have extrapolated $T_C(x)\to 0$ from $T>3K$ at
$T_C=0$ for $x\approx 0.65$. However, recent studies performed at
lower temperature revealed that $T_C(x)$ does not tends to zero at
the previously reported value because its negative curvature
changes to slightly positive at $x^*\approx 0.63$. Consequently
the $T_C\to 0$ extrapolation was proposed at $x_{cr}\approx
0.85$.\cite{13sces04} Since the lowest transition temperature at
T$_C=0.25K$ was observed for $x=0.80$, CePd$_{1-x}$Rh$_x$ became
one of the few systems, and the only F- one \cite{14Ser01}, in
which T$_C(x)$ was traced within more than one decade in
temperature. These properties make CePd$_{1-x}$Rh$_x$ an ideal
candidate for the study of F quantum phase transitions.

\section{Experimental details and results}

A series of nine samples was prepared in the range $0.6\leq x \leq 1$ by arc melting the appropriate amount of pure elements under argon atmosphere. In order to ensure homogeneity, the resulting buttons were flipped over and remelted several times. For a further improvement of the homogeneity the buttons were melted again in a high frequency levitation crucible. At the end of this process the mass loss was found to be negligible (generally below 0.1\%). The samples were then annealed for 7 days at 700$^{\circ}$\,C in dynamic vacuum. Powder XR diffraction of crushed samples confirmed the orthorhombic CrB structure \cite{15Parte} and did not reveal sign of secondary phases. Since the Bragg peaks of the as crushed samples were rather broad, the powder used for the XR measurements was annealed at 700$^{\circ}$\,C for 24 hrs in dynamic vacuum, leading to much narrower peaks. Using these sharp diffraction patterns we could determine precise values of the lattice parameters.

\begin{figure}
\begin{center}
\includegraphics[angle=0,width= 0.70 \textwidth] {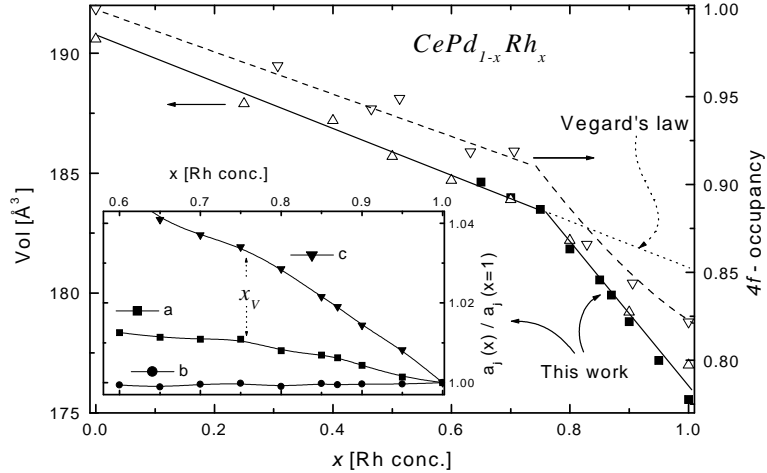}
\end{center}
\caption{Concentration dependence of the unit cell volume compared
with data from Ref \cite{11CePdRh}. A clear deviation from
Vagard's law occurs at $x_V\approx 0.75$ at the onset of the
valence instability, where also $L_{III}-XAS$ spectroscopy
indicates a stronger decrease of 4$f$ occupancy \cite{11CePdRh}.
Inset: normalized variation of each lattice parameter (a$_j$)
showing the crystalline anisotropy of this system.}
\label{F1lattpar}
\end{figure}

A standard SQUID magnetometer served for the determination of the
magnetization from 2\,K up to room temperature in a 0.1\,T magnetic field. The
ac-susceptibility was measured using the mutual inductance technique, with a
lock-in amplifier as detector working between 0.1 and 12.8 kHz with an
excitation amplitude of $\approx 10 \mu T$ in the $0.5\leq T \leq 6$\,K range. For
lower temperatures (i.e. 18\,mK $ \leq T \leq 4$\,K) a dilution refrigerator was
used with the same excitation amplitude and frequencies between 13\,Hz and
1113\,Hz. The electrical resistivity was measured using a four probe dc-method
in the temperature range from 0.5\,K up to room temperature. Specific heat $C_P(T)$ 
measurements on samples of about 1g were performed in a semi-adiabatic
calorimeter at temperatures ranging from 0.5\,K up to 20\,K, using a heat pulse
technique. Thermal expansion was measured along three perpendicular directions of polycrystals of rectangular shape $\alpha_i$ with the aid of a ultra-high resolution cpacitive dilatometer. 
The volume expansion coefficient $\beta(T)=1/V dV/dT$ is obtained from the sum of the linear expansion coefficients $\beta =\alpha_1 + \alpha_2 + \alpha_3$.

\subsection{High-temperature results}

\begin{figure}
\begin{center}
\includegraphics[angle=0,width= 0.70\textwidth] {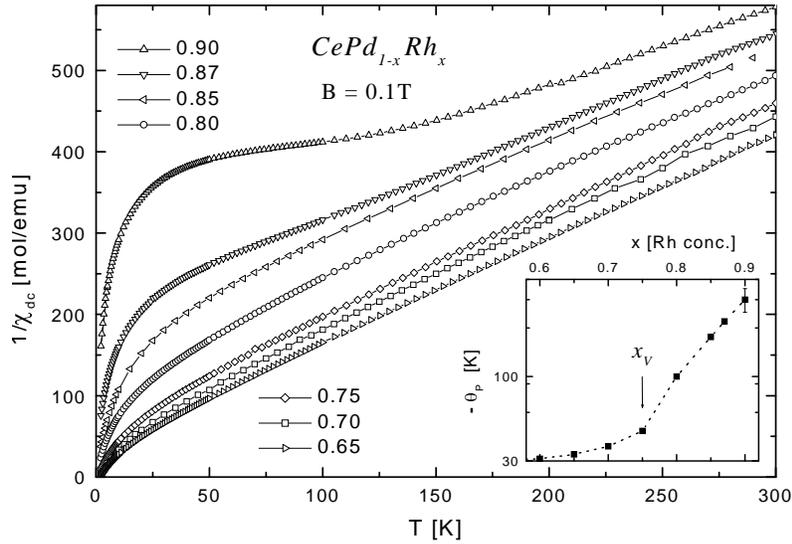}
\end{center}
\caption{Inverse dc-susceptibility as a function of temperature
showing the increasing hybridization effects for $x>0.75$. Inset:
concentration dependence of Curie-Weiss temperature $\theta_p$ in
logarithmic scale, extrapolated from $T\geq 100$\,K ($\geq 200$\,K for
$x=0.90$).} \label{F2invChi}
\end{figure}
As mentioned before, the cell volume decreases continuously with
increasing Rh content and the new measurements on intermediate
concentrations allow to better establish the deviation form
Vegard's law at $x_V=0.75$, as shown in Fig.1. Because of the
anisotropic crystalline structure of CePd$_{1-x}$Rh$_x$, we have
analyzed the $x$ dependence on each crystalline parameter. In the
inset of Fig.1 one observes that the change of slope at $x_V=0.75$
occurs on "$a$" and "$c$" directions, while "$b$" remains
practically unchanged. In order to confirm that such a deviation
from Vegard's law corresponds to the onset of the valence
instability, we include in Fig.1 the $L_{III}-XAS$ results from
Ref. \cite{11CePdRh}. These results coincide in the $x_V$
determination as the concentration for the onset of the 4$f$
occupancy decrease.

DC-susceptibility ($\chi_{dc}$), measured under an applied
magnetic field of $B$=0.1T, were carried out on samples within the
$0.6\leq x \leq 1$ concentration range. The temperature dependence
between 2K and room temperature is depicted in Fig.2 as
$\chi_{dc}^{-1}$. The high temperature (T$>$100K) dependence of
samples $0.65\leq x \leq 0.87$ can be described by a typical
Curie-Weiss (CW) law: $\chi=C_c/(T-\theta_P)$, with a negative
$\theta_P$ despite its F-GS   (see inset of Fig.2). Beyond the
critical concentration, it shows the features of a
growing valence instability which dominates the signal CeRh (c.f.
$x=1$), with a broad maximum centered at $\approx 240$\,K (not
shown). As it occurs in systems undergoing a magnetic to mixed
valence transition, the molecular field (represented by
$\theta_P$) is overcome by the Kondo effect. For that case a
$\chi=C_c/(T+2T_K)$ expression was proposed \cite{16Krishnam}
where $T_K$ is the Kondo temperature reflecting the $J_K$ increase
since $T_K \propto 1/\delta J_K$. \cite{6Doniach}
\begin{figure}
\begin{center}
\includegraphics[angle=0,width= 0.5\textwidth] {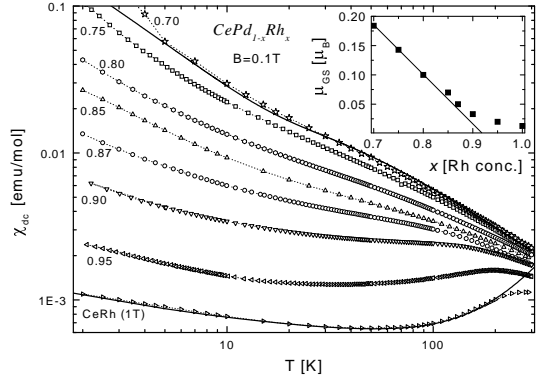}
\end{center}
\caption{Temperature dependence of dc-susceptibility in a double
logarithmic representation, showing the effect of hybridization on
the excited CEF levels at high temperature and the weakening of
the effective moments at low temperature. Continuous curves are
fits for $x=0.70$ and $x=1$ data using eq(1) and eq.(2)
respectively.} \label{F3chiLoglog}
\end{figure}

The crystalline electrical field (CEF) effect can be observed in
$x=0.70$ with $\chi_{dc}(T)$ plotted as a function of temperature
in Fig.3, in a double logarithmic representation. To evaluate the
energy of the first excited level ($\Delta_{CF}$) we used a
simplified formula applicable to the low temperature range (i.e.
$T<\Delta_{CF}$):
\begin{equation}
\chi(T)_{CF}=\mu^2_{GS}+\mu_{CF}^2 \times exp(-\Delta_{CF}/T)/(Z \times T)
\end{equation}
where $\mu_{GS}$ ($\mu_{CF}$) is the ground (excited) state
effective moment and $Z$ the partition function. Using this simple
equation we obtain $\Delta_{CF}\approx 55K$ and the respective
effective moments: $\mu_{GS}\approx 0.18 \mu_B$ and
$\mu_{CF}\approx 0.34\mu_B$ (see continuous curve on $x=0.70$ in
Fig.3). The strong increase of the $\chi_{dc}$ below 4K is due to
the onset of ferromagnetic correlations. The progressive weakening
of $\mu_{GS}(x)$ evaluated around 2\,K is depicted in the inset of
Fig.3. The $\mu_{GS}$ points between $0.70<x<0.80$ extrapolate to
zero at $x \approx 0.90$, whereas the deviation from such
extrapolation for $x\geq 0.90$ can be related to remnant low
energy excitations observed in $C_m/T$ measurements to be
discussed below.

On the non-magnetic limit, CeRh shows the typical behavior for a
mixed valence (MV) compound plus a power law increase at low
temperature:
\begin{equation}
\chi(T)=\chi_0+b T^2+c/T^q
\end{equation}

The Pauli-type contribution is $\chi_0=0.46 \times 10^{-3}$\,emu/mol and the $b T^2$
dependence reflects the presence of spin fluctuations, in this case with $b =
0.12 \times 10^{-3}$ emu/molK$^2$. The increase of $\chi(T)$ below about 30K indicates a
reminiscence of magnetic moments even at the CeRh limit. This magnetic
contribution cannot be attributed to magnetic impurities only because its
temperature dependence: $c/T^q\propto T^{-0.45}$ does not correspond to a
Curie-Weiss law. $M$ vs $B$ measurements at T=2\,K and up to B=5T (not shown)
indicate that only a small magnetic contribution, which saturates at
$B\approx 3$\,T is due to impurities. For this compound the Wilson ratio,
$\chi_0 /\gamma_0=0.036$\,emuK/mol corresponds to the expected value for an MV
(i.e. six fold degenerated) ground state \cite{4Ser91}. Once normalized:
$R=(\pi k_B/\mu_{eff})^2$$\chi_0 /\gamma_0=1+1/2J=1.2$, since for Ce-MV
systems $J=5/2$. \cite{Besnus}

\begin{figure}
\begin{center}
\includegraphics[angle=0,width= 0.5\textwidth] {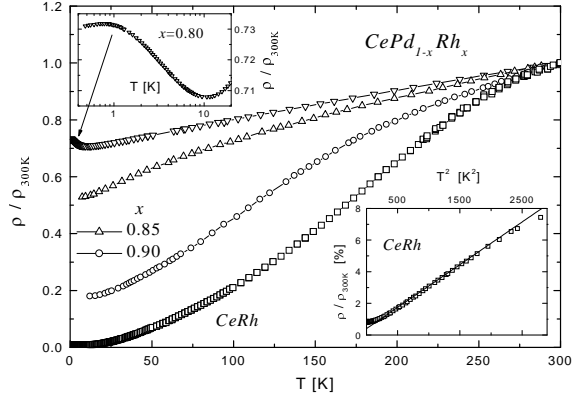}
\end{center}
\caption{Electrical resistivity normalized at room temperature
showing the rapid decrease of the low temperature scattering above
$x_{cr}$ and the increase of hybridization effect with the valence
instability. Upper inset: detail of the merging magnetic
scattering in the $x=0.80$ sample. Lower inset: detail of the
$\rho\propto T^2$ dependence of CeRh up to 40K and the slight
upturn below 15K.} \label{F4Resist}
\end{figure}

The electrical resistivity ($\rho$) measured up to room
temperature of the samples within the $0.80< x < 1$ Rh
concentration is displayed in Fig.4. There, our results on the
$x=0.80$ and $1$ samples are compared with those of $x=0.85$ and
$0.90$ (after Ref.\cite{11CePdRh}) once normalized to their
respective values at room temperature. Two features characterize
$\rho(T)$, one is the rapid increase of the hybridization effect
on the high temperature electronic scattering and the other the
drastic drop of the residual resistivity ($\rho_0$) as $x\to 1$.
The former is concomitant with the increase of $\theta_P$
($\propto T_K$) in that concentration range. In the case of
$x=0.80$ a $\rho(T)$ upturn occurs below 10K (see upper inset in
Fig.4). Such an electronic scattering arises nearly two decades
above the $\chi'_{ac}$ cusp and thermal expansion minimum (at
$T=0.25K$) and is probably due to magnetic correlations related to
that transition. The lower inset in Fig.4 shows the analyzed
$\Delta \rho \propto T^2$ dependence up to $\approx 40$\,K. The
weak upturn observed at low temperature will be analyzed in the
context of other low temperature properties.

\subsection{Low temperature results}

In order to determine $T_C(x)$ below the lowest value previously
reported ($T_C=3$\,K \cite{12CePdNiRh}), we have performed two
sets of $\chi_{ac}$ measurements, one down to 0.5K in a He$^3$
cryostat and the other down to 20mK in a dilution refrigerator.
The F-phase boundary was traced following the temperature of the
maximum of the inductive signal $\chi '_{ac}(T)$ in the $0.60\leq
x \leq 0.90$ samples. These results are presented in Fig.5, after
normalizing the measured signal with their values at 5K and the
respective maxima. The sharpness of the anomaly (not depending on
$x$) also indicates that eventual atomic disorder introduced by
Pd/Rh substitution does not affect the F- transition, which keeps
its shape down to the critical concentration ($x_{cr}$).
\begin{figure}
\begin{center}
\includegraphics[angle=0,width= 0.5\textwidth] {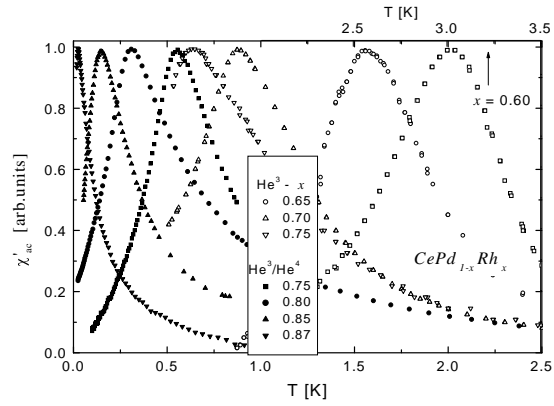}
\end{center}
\caption{Inductive signal of ac-susceptibility, $\chi'_{ac}$,
normalized at their respective maximum values. For clarity, the
temperature dependence of sample $x=0.60$ (only) is depicted on
the upper abscissa of the figure.} \label{F5Chi1}
\end{figure}

The electronic contribution to the specific heat ($C_{el}$) was
obtained by subtracting the phonon contribution ($C_{ph}$) to
measured values ($C_P$) as: $C_{el}=C_P - C_{ph}$. The
stoichiometric compound LaRh was taken as the reference for the
$C_{ph}$ determination. Fig.6 collects the $C_{el}/T$ results.
Though the specific heat magnetic anomaly ($C_m$) dominates
$C_{el}$ at the F-phase, the experimental data indicate that with
increasing $x$ a Fermi liquid (FL) contribution, $\gamma(x)$,
arising from the MV component, becomes important and then
$C_{el}/T$ has to be accounted as $C_{el}/T = C_m/T + \gamma $ for
high Rh concentration. It can be seen in Fig.6 that sample $x=0.50$ still shows a sharp transition 
that broadens for $x\geq 0.60$ (i.e. for $x\geq x^*$). However, the maximum of
$C_{el}/T$ stops decreasing and becomes nearly constant for the $x=0.60, 0.65$ and $0.70$ samples, like in
other Ce-systems tuned to their quantum critical points (QCP).\cite{Pablo} Beyond $x_{cr}$ the $\gamma$
component tends to dominate the total contribution and, for
samples $x=0.90, 0.95$ and $1$, the FL contribution is found to be
$\gamma = 0.047, 0.028$ and $0.017$\,J/molK$^2$ respectively. This
type of contribution was observed in many Ce-systems at the onset
of their valence instabilities.\cite{8Ser95}
\begin{figure}
\begin{center}
\includegraphics[angle=0,width= 0.5\textwidth] {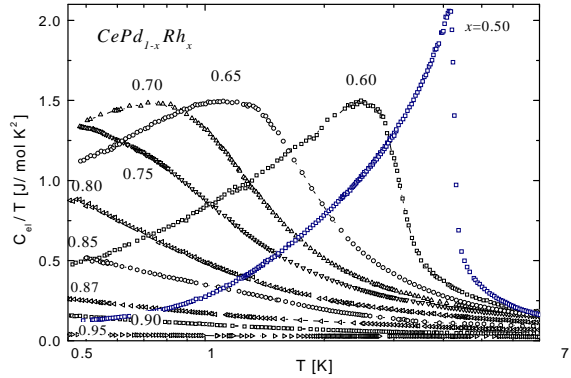}
\end{center}
\caption{Total electronic contribution to the specific heat divided temperature as a function of
temperature in a logarithmic scale for samples around the critical region.}
\label{F6CmVsT}
\end{figure}
\begin{figure}
\begin{center}
\includegraphics[angle=0,width= 0.5\textwidth] {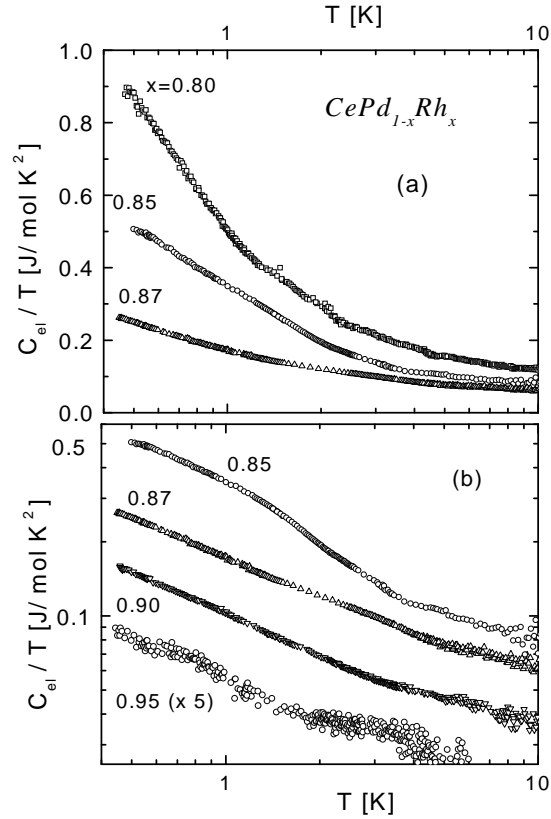}
\end{center}
\caption{a) Specific heat divided temperature showing the
logarithmic dependence on T at low temperature on the critical
concentration. b) Double logarithmic representation showing the
power law temperature dependence beyond the critical
concentration. For sample $x=0.95$, the $\gamma$=0.028J/molK$^2$
contribution is subtracted (see the text). Notice that data from
samples $x=0.85$ and $0.87$ are also included in panels b) and a)
respectively for comparison.} \label{F7Log&Arrenius}
\end{figure}
A detailed analysis on the temperature dependence of $C_{el}(T)/T$
around $x_{cr}$ is made in Fig.7. There, a logarithmic temperature
dependence is observed for $x=0.80$ and $0.85$ samples (see
Fig.7a), whereas $x=0.87, 0.90$ and $0.95$ show a power law T
dependence as proved by the double logarithmic representation in
Fig.7b. These results are described by a $C_m/T= A T^{-q}$ formula
with the exponent $q=0.54\pm 0.01$ and $A= 0.173, 0.102$ and
$0.012$\,J/mol\,K$^{2-q}$ respectively. In the case of sample
$x=0.95$, the $\gamma(x)$ contribution was subtracted since
$C_m(T)/T$ becomes very small at this Rh concentration.

\begin{figure}
\begin{center}
\includegraphics[angle=0,width= 0.5\textwidth] {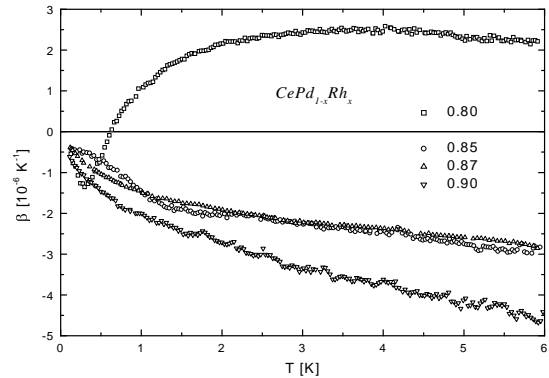}
\end{center}
\caption{Thermal expansion measurements showing $T_C$ at 0.25K and the sudden change of
sign at $x=0.85$.} 
\label{F8allbetas}
\end{figure}
Thermal expansion measurements performed down to $T\approx
0.1K$ on samples $0.80 \leq x \leq 0.90$ are shown in  Fig.8. The
magnetic transition in the $x=0.80$ alloy is clearly identified by
a minimum at 0.25K, in coincidence with the $\chi'(T)$
measurements. Noteworthy is the change of sign of $\beta$ at
$x=0.85$ in coincidence with the already mentioned changes in
other experimental parameters. Nevertheless, it has to be
mentioned that in this sample there is still a positive
contribution observed in one of the measured directions. The
anisotropic expansion of this system has to be related with the
intrinsic crystalline anisotropy. Detailed analysis of the non-Fermi liquid behavior of $\beta(T)$ will be presented elsewhere and compared with that of C$_m(T)$.

\section{Discussion}

The competition between the vanishing F- order and the growing
Kondo screening with Rh concentration is visualized in the
magnetic phase diagram in Fig.9. There one appreciates that,
instead of a continuous negative curvature in $T_C(x)$, an
inflection in the phase boundary occurs at $x^*=0.65$. The
characteristic of this concentration is that $T_K(x)$ (considered
proportional to $\theta_P(x)$\cite{16Krishnam}) starts to increase
as shown on the right side of Fig.9 in a logarithmic scale. Though
Doniachs model includes both magnetic and Kondo interactions, it
predicts a vanishing $T_C(x)$ without any possible change of
curvature. In order to discuss such a case we have to take into
account that two critical behaviors occur at similar concentration
as it is indicated by high and low temperature measurements. One
concerns the valence instability signed by the deviation form
Vegard's law and $4f$ occupation reduction at $x\geq x_v$ and the
other the magnetic critical point at $x_{cr}\approx 0.87$. Though
this peculiar situation was already observed in a few Ce-systems,
\cite{8Ser95} it was never studied in detail before. In fact, the
present investigation demonstrates that the previous $T_C\to 0$
extrapolation to $x\approx 0.65$ \cite{11CePdRh} did not account
for the change of curvature occurring at $x^*=0.60$ that
originates an extended "tail" at low temperatures. Besides the
change of curvature there are other clear indications for
significant modifications in the GS properties between $x^*$ and
$x_{cr}$. 

As mentioned before, the well defined $C_m/T(T_C)$ jump at $x=0.50$ broadens for $x=0.60$, keeping a similar relative width and the same maximum value up to $x=0.70$. Since no sudden changes in the atomic order are expected between $x=0.50$ and 0.60, this modification confirms the changes occuring  around $x^*$. 
At higher Rh concentration (c.f. 0.80) the $C_m/T(T)$ dependence becomes logarithmic in 
agreement with theoretical predictions for 3-dimensional itinerant
ferromagnets.\cite{Millis} Coincidentally, the $\chi_{ac}$ maximum
starts show frequency ($w$) dependence for $x\geq 0.70$. These
features are usually associated to atomic disorder in diluted
systems. However, due to the {\it lattice} character of the
magnetic atoms and the similar atomic size of Ce ligands (c.f. Pd
and Rh) a sudden appearance of a spin glass like behavior between
$x=0.60$ and $0.70$ is unlikely. Nevertheless, in an ample concept
of disorder, the presence of random distribution of interactions
competing in certain ranges of energy have to be taken into
account. Magnetic ($J_R$) and Kondo ($J_K$) interactions, powered
by respective Pd and Rh Ce-neighbors with random spacial
distribution, are expected to build up an inhomogeneous pattern
for those parameters. Since in such a scenario no more long range
F-order is expected, for $x>x^*$ we will identify the line of
$\chi_{ac}$ maxima as $T_C^*(x)$.

It was shown that random hybridization ($\propto J_K(x)$) leads to
a finite probability of very small $T_K$ values ($T_K^L$) and to
non-Fermi-liquid behavior.\cite{Miranda} This may explain that a
magnetic character of the GS may coexist with a strong increase of
$T_K(x>x_V)$. On the other hand, spin disordered systems close to
their QCP were described as Grifftihs phases due to the possible
formation of non percolating magnetic clusters.\cite{20CastroN}
Since this scenario predicts a power law $C_m(T)$ dependence, for
our case it only applies to the $x\geq x_{cr}$ region. A more
realistic description can be done accounting for an inhomogeneous
distribution of each Ce environment, where $J_R(x)$ and $J_K(x)$
act simultaneously.\cite{Lobos}

The strong anisotropy of this system is also expected to play a
role in the coexistence between the magnetic GS and the strongly
hybridized excited CEF levels wave functions. The orthogonality of the CEF levels
wave functions leads to the possibility of different intensities
between local and conduction spin couplings due to their different
mixing matrix elements. Such a situation becomes sensitive when
$T_K \approx \Delta_{CF}$ since the broadening of the excited CEF
levels increase their influence on the low energy range.
\begin{figure}
\begin{center}
\includegraphics[angle=0,width= 0.5\textwidth] {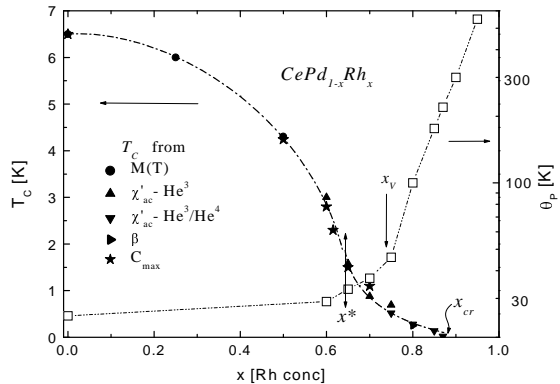}
\end{center}
\caption{Magnetic phase diagram along the full concentration range, extracted from
magnetic ($M, \chi'_{ac}$) and thermal ($\beta, C_m/T|_{max}$) properties. $x^*$ marks
the inflection of $T_C(x)$ curvature, $x_V$ the onset of the valence instability and
$x_{cr}$ the critical point. The concentration dependence of the Curie-Weiss temperature
is included for comparison in a logarithmic scale (on the right axis) to show its rapid
increase for $x>x_v$. $\Delta_{CF}$ indicates the estimated first CEF splitting. Dash-dot
lines are guides to the eye.} \label{F9TcVsRh}
\end{figure}

The actual hybridization effect (given by $T_K^L$) on the low-lying levels can be evaluated by means of a low temperature
property like the magnetic entropy ($S_m(x,T)$).
Taking profit that $S_m\propto 1/T_K$ \cite{18Schotte}, one can
obtain $T_K^L(x)$ from the measured $S_m$ at a fixed temperature.
In this case, we have extracted $S_m$ at $T=8K$ from the $S_m(T)$
data shown in Fig.10 and computed the $RLn2/S_m(x)$ values as
depicted in the inset of Fig.10. Though on the Rh rich side its
tendency is similar to the extracted form $\theta_P(x)$, a clear
deviation between those parameters is observed between $x_V$ and
$x_{cr}$ since the GS and the CEF excited levels are differently
affected. Further increase of $J_K(x)$ leads to the collapse of
the CEF effect and the sixfold degenerated Hund's rule GS takes
over at the MV limit.

\begin{figure}
\begin{center}
\includegraphics[angle=0,width= 0.5 \textwidth] {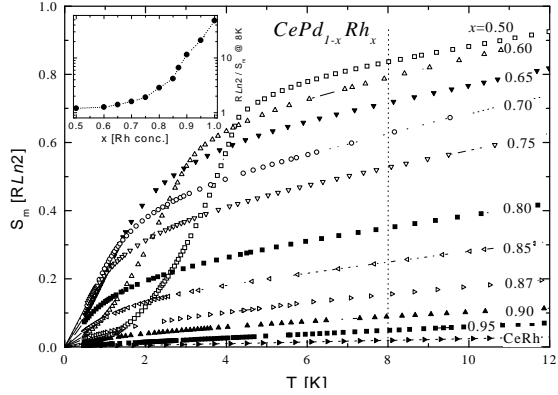}
\end{center}
\caption{Entropy gain on temperature for all studied samples, normalized to $RLn2$.
Inset: inverse of entropy at fixed temperature (8K) as a function of concentration (see
the text).} \label{F9Entrop}
\end{figure}

The change of sign in the thermal expansion indicates that the volume behaves
completely different on both sides of the critical concentration. Starting
from $T=0$, sample $x=0.80$ first contracts to undergo the magnetic
transition, but then rapidly expands. On the contrary, for $x\geq 0.85$ (i.e.
$x_{cr}$), the volume continuously contracts (at least up to 6K). Even being
anisotropic, the negative sign of $\beta(T)$ for $x\geq 0.85$ is quite
surprising because in that concentration range the MV regime becomes dominant.
In fact, well known Ce-MV compounds show positive $\beta(T)$ coefficient (e.g.
CeSn$_3$ \cite{CeSn3}) because the main energy scale, related to $T_K$,
increases under pressure. However, since this observation corresponds to the
low temperature range (i.e. up to 6K $\ll T_K$) it only involves the low
energy excitations of the system, where a diverging $C_m/T$ component is still
present. Taking profit that the linear thermal expansion of each sample was
measured in three perpendicular directions, further information about
anisotropic effects of this system can be extracted. Notably, only in one
direction $\alpha (T)$ is positive (arbitrarily label as $\alpha_1$) and
becomes zero at $x=0.87$, whereas the other ($\alpha_2$ and $\alpha_3$) show
an increasing negative temperature dependence. Further measurements on single
crystals are certainly required to relate this anisotropic behavior to the
actual crystalline axis, see Ref.\cite{21deppe}
\begin{figure}
\begin{center}
\includegraphics[angle=0,width= 0.5\textwidth] {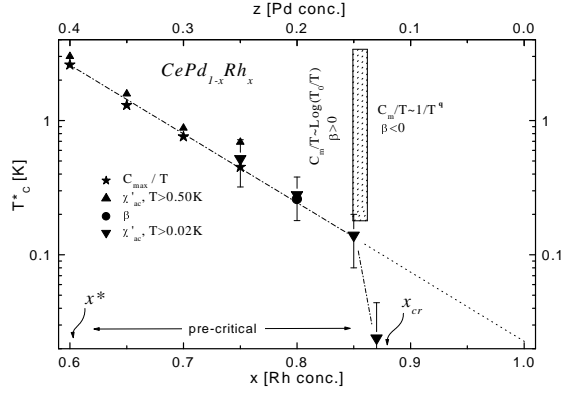}
\end{center}
\caption{$T_C^*(x)$ dependence at the pre-critical concentration showing its
logarithmic decrease and its sudden disappearance of the $\chi'_{ac}(T)$ cusp.
The change of thermal properties is indicated around the critical
concentration.} \label{11HT&LTPhsDiag}
\end{figure}

In Fig.11 we show a detailed $T_C^*(x)$ dependence obtained from
the different techniques, notably $\chi'_{ac}$ at very low
temperature, which are depicted in a logarithmic scale. The
positive curved "tail", already presented in the phase diagram of
Fig.9, can be described by a simple function:
\begin{equation}
T_C^*(z) = 0.022\,K \times e^{12z}
\end{equation}
where $z=1-x$ corresponds to Pd concentration. This permutation of
the concentration parameters dependence is not a minor detail
since it indicates that the $T_C^*(z)$ evolution refers to a
$smeared$ phase transition \cite{Sknep} even beyond the CeRh
limit, in agreement with the remnant contribution of $C_m/T$ on
that side of the phase diagram.  Since this description is valid
for $z>0.13$, it seems that some type of {\it percolation} between
magnetic $rare$ regions \cite{1Voijta} is needed to produce the
cusp detected by $\chi_{ac}$. Such a {\it percolation} threshold
is in agreement with the onset of the power law dependence of
$C_m/T$ since, as mentioned before, the formation of Griffiths
phases requires the existence of non-percolating clusters, that
occurs for $x>0.87$. The fact that the $T_C^*$ variation can be
described by such a simple $z$ dependence suggests that the
physics at the {\it pre-critical} region is governed by a critical
behavior lying beyond the CeRh limit. This pattern implies a split
of the phase diagram description at $x\approx x^*$, which seems to
present two distinct regions, represented by the $T_C$ boundary
and the $T_C^*$ line respectively. On the canonical F- (Pd rich
side c.f. $0\leq x \leq x^*$) the $T_C(x)$ phase boundary (between
$6K \geq T_C > 2K$) points to the former \cite{12CePdNiRh}
critical point at $x\approx 0.65$. Beyond $x^*$ (i.e. below about
2K) another type of criticality dominates the scenario according
to the drastic changes in the low temperature properties. The
question arises whether this threshold is governed by the
concentration (i.e. chemical potential) or by the exhausting
thermal fluctuations overcome by quantum fluctuations.

Such a $thermal$ threshold is not an arbitrary proposition since
other well known AF-Ce systems show similar changes in their
magnetic phase boundaries at that temperature (c.f. thermal
energy) even close to a first order transition, like:
CeIn$_{3-x}$Sn$_x$ with $x^*\approx 0.40$, where $T_N=2$\,K,
\cite{Pablo} and CeCu$_2$(Si$_{1-z}$Ge$_z$)$_2$ with $z^*=0.25$\,K
and also $T_N=2$\,K \cite{Oliver}. Notice that parameters "$x$"
and "$z"$ are used like in the present context, i.e. $x\to 1$
($z\to 0$) is the non-magnetic side.

Going back to Rh concentration $x$ as the control parameter, another important
observation is the frequency ($w$) dependence of the $\chi'_{ac}$ cusp. Such a
dependence is only observed within $0.70\leq x \leq 0.87$ and the question
arises whether it is due to a canonical spin-glass behavior or to fluctuations
related to the QCP. Since a $C_m/T \propto -log(T/T_0)$ dependence is
predicted for a Non-Fermi-Liquid behavior (instead of the $\propto 1/T^2$ for
a spin-glass), quantum critical fluctuations may be the responsible such a
$\chi'_{ac}(w)$ dependence.\cite{Grempel} Nevertheless, other hallmarks of
spin-glass GS, like the difference between zero-field and field-cooling or the
time dependence of $\chi'_{ac}$ response under field suppression have to be
explored.

\section{conclusions}

The phase diagram of F- CePd$_{1-x}$Rh$_x$ was traced between $T_{C,x=0}=6$\,K
and $T^*_{C,x=0.87}\approx 20$\,mK. Three characteristic concentrations were
identified: $x^*\approx 0.65$; $x_V\approx 0.75$ and $x_{cr} \approx 0.87$.
The former marks the cross over from a classical phase boundary (with a
negative curvature) into a positive curved "tail" which ends at
$x_{cr}=0.87$. In that region a frequency dependence of the $\chi'_{ac}$ cusp
is also detected.

Independently from this magnetic GS, a deviation from Vegard's law
occurs at $x_V$, indicating the onset of the valence instability.
Beyond that concentration $T_K(x)$ increases exponentially. Unlike
typical AF systems, where $x_V > x_{cr}$, this ferromagnet has its
critical concentration ($x_{cr}\approx 0.87$) above the onset of
the valence instability and therefore $x_V<x_{cr}$. Such a
characteristic inhibits the appearance of HF behavior because at
that concentration $J_K$ is already largely developed, leading to
the broadening of the quasi-particles band with the consequent
reduction of their effective mass. Besides this $\gamma(x)$
decrease, there is a remanence of low laying magnetic excitations
accounted by the $C_m/T\propto 1/T^q$ contribution. Such a
coexistence of magnetism with highly hybridized quasi-particles
occurs in an inhomogeneous distribution of $J_R$ and $J_K$
couplings produced by the random distribution of Pd and Rh
Ce-ligands.

The different magnetic behavior observed between high and low
temperature properties of this system is related to its strong
anisotropic character. Such anisotropy is clearly observed in the
crystalline parameters and in thermal expansion measurements.
Therefore a different degree of hybridization occurs between the
ground and the excited CEF levels.

The outstanding findings around $x_{cr}$ are the change temperature dependence of $C_m/T$, form
logarithmic to power law coincident with the change of sign of $\beta$. Notably, around that concentration $C_m/T$ and $\chi_{dc}$ coincide in their $T^{-q}$ temperature dependence. Though in the pre-critical ($x^* < x < x_{cr}$) region, the logarithmic $C_m/T(T)$ dependence is in agreement with the theoretical predictions for 3-dimensional itinerant ferromagnets,
its transformation to a power law beyond $x_{cr}$ indicates that a
sort of percolation between "rare" regions can play an important
role in this behavior. The simultaneous effect of ferromagnetic
and valence fluctuations may result in a more complex scenario for
the electronic scattering, which is not accounted in the model.

Finally, we have shown that a better understanding of the phase
diagram can be achieved by splitting it into two regions. One with
the long range F-order extrapolating $T_C(x)\to 0$ at $\approx 0.65$ and the
other (below about 2K) extrapolating $T_C^*(x)$ beyond the CeRh
limit. Altogether, CePd$_{1-x}$Rh$_x$ has proved to be an
exemplary system for studying the influence of doping on the
quantum criticality in a Ferromagnetic environment, covering two
and a half decades of temperature.

\section*{Acknowledgments}

This work was partially supported by F. Antorchas - DAAD cooperation program
(proj. \# 14248/7). J.G.S. is member of the CONICET and Instituto Balseiro of
Argentina.



\end{document}